\documentstyle [12pt,epsf]{article}

\newcommand{\s}{\sigma}
\newcommand{\ep}{\epsilon}

\newcommand{\La}{\Lambda}
\newcommand{\g}{\gamma}
\newcommand{\p}{\partial}
\newcommand{\m}{\mu}
\newcommand{\n}{\nu}
\newcommand{\al}{\alpha}
\newcommand{\ti}{\tilde}
\newcommand{\om}{\omega}
\newcommand{\be}{\begin{equation}}
\newcommand{\ee}{\end{equation}}

\begin{document}

\title{\bf
An Example of Higher Weight Superpotential Interaction in the
Heterotic String on Orbifolds
}
\author{
{\bf D.Erdenebayar }\\
\normalsize International Centre for Theoretical Physics, Trieste
34100, {\bf Italy}\\
{\normalsize and}\\
\normalsize The Institute of Physics and Technology, Ulaanbaatar,
{\bf Mongolia}\thanks{Permanent Address} }
\maketitle

\begin{abstract}
An explicit orbifold example of the non-zero correlation functions
related to the additional contribution to the induced mass term for
Higgs particles at low energies is given. We verify that they form
finite dimensional representations of the target space modular
transformation $SL_2(Z)$. This action of the modular group is shown
to be consistent with its action on the fixed points set defining the
twisted fields.
\end{abstract}
\section{Introduction}
\newcommand{\K}{K\"{a}hler\* }

Minimal N=1 supergravity Lagrangian involving two-derivative terms
is described by three functions of chiral and anti-chiral superfields
$(\Phi,\bar\Phi)$, which live on a \K space. These functions are \K
potential $K(\Phi,\bar\Phi)$, chiral superpotential $W(\Phi)$ and the
gauge coupling $f(\Phi)$. \K potential determines the metric on the \K
space and therefore the kinetic term of the fields $\Phi$, whereas the
superpotential $W$ determines the Yukawa couplings. The fields
$\Phi$ in general contain gauge-charged as well as gauge neutral
fields. Among the latter there can be some fields that correspond to
flat direction of the potential (i.e. ${\p V\over\p\Phi}=0$ identically).
Such fields are called moduli fields as they can acquire arbitrary
constant vacuum expectation values.

A general feature of the N=1 supergravity models coming from
superstring theories is the presence of such moduli fields. They
correspond to exactly marginal deformations of the underlying
superconformal field theory describing the internal manifold. Besides
these moduli fields, which depend on the details of the internal
theory, there is a universal field $S$ called dilaton, which is also a
flat direction of the potential. In fact the constant v.e.v. of dilaton
field determines the string coupling constant. The tree level gauge
function $f$ turns out to be only a function of $S$: the real part of $S$
defines the usual gauge coupling constant and the imaginary part
determines the coupling of $F\ti F$ term. \K potential $K$ and
superpotential $W$ on the other hand are model dependent.

In the case of (2,2) compactification of the Heterotic superstring the
gauge group is $E_{6}\times E_{8}$ and the matter fields transform as
\boldmath $27$ or $\overline{27}$ \unboldmath under $E_{6}$ and
they are in one-to-one correspondence with the moduli. The \K
potential has the following power in expansion in the matter fields: $$
K=G+A^{\al}A^{\dot{\al}}Z^{(1,1)}_{\al \dot{\al}}
+B^{\n}B^{\dot{\n}}Z^{(1,2)}_{\n \dot{\n}} +(A^{\al}B^{\n}H_{\al
\n}+c.c.)+\dots
$$
where $A$ and $B$ refer to \boldmath $27$'s and $\overline{27}$'s
\unboldmath respectively.
The functions $Z^{(1,1)}_{\al \dot{\al}}$ and $Z^{(1,2)}_{\n \dot{\n}}$
determine the kinetic terms and have been studied extensively in the
literature \cite{geometry} and were shown to be related to Yukawa
couplings via special geometry relations. $H$ function on the other
hand gives rise to mass term (called $\m$ term) once supersymmetry
is broken.

The most convenient way of computing the $H$ function at the tree
level is to consider the four-scalar scattering amplitude ${\cal A}
(A^{\al},B^{\n},M^{\bar m},M^{\bar n})$
\begin{figure}
\setlength{\unitlength}{1cm}
\begin{picture}(15,5.7)(0,0.3)
\put(1.5,4){\line(1,1){1}} \put(1.5,4){\line(1,-1){1}}
\put(1.5,4){\line(-1,1){1}} \put(1.5,4){\line(-1,-1){1}}
\put(0.1,4.8){$\bar m$}\put(0.1,3.0){$\bar n$}
\put(2.6,4.8){$A^{\al}$}\put(2.6,3.0){$B^{\n}$} \put(3.3,3.9){+}
\put(5.1,4){\line(-1,1){1}} \put(5.1,4){\line(-1,-1){1}}
\put(6.6,4){\line(1,1){1}} \put(6.6,4){\line(1,-1){1}}
\put(5.1,4){\line(1,0){1.5}}
\put(3.7,4.8){$\bar m$}\put(3.7,3.0){$\bar n$}
\put(7.7,4.8){$A^{\al}$}\put(7.7,3.0){$B^{\n}$}
\put(5.2,4.1){$k$}\put(6.3,4.1){$\bar k$} \put(8.4,3.9){+}
\put(10.2,4){\line(-1,1){1}} \put(10.2,4){\line(-1,-1){1}}
\put(11.7,4){\line(1,1){1}} \put(11.7,4){\line(1,-1){1}}
\put(10.2,4){\line(1,0){1.5}}
\put(8.8,4.8){$\bar m$}\put(8.8,3.0){$\bar n$}
\put(12.8,4.8){$A^{\al}$}\put(12.8,3.0){$B^{\n}$}
\put(10.3,4.1){$h^{\bar s}$}\put(11.4,4.1){$h^s$}
\put(1,1.8){$\underbrace{(p_Ap_B)\p_{\bar m}\p_{\bar
n}H_{\al\n}\hspace{1cm}
(p_Ap_B)\Gamma_{\bar m\bar n}^{\bar k}\p_{\bar
k}H_{\al\n}}_{(p_Ap_B) \nabla_{\bar m}\p_{\bar n}H_{\al\n}}
\hspace{1cm}(p_Ap_B)T_{\bar m\bar n\bar s}G^{\bar s s}W{\al\n s}$}
\put(1,0.5){where $h^s$ is the auxiliary component of gauge singlet
$s$.} \end{picture}
\end{figure}
involving \boldmath $27$, $\overline{27}$ \unboldmath and two
antimoduli. In the leading, quadratic order in momenta, there exist in
general two contributions to the process. The first, coming from the
standard D-term interactions in the effective Lagrangian, gives the
Riemann tensor of \K geometry, $R_{\al \bar m \n \bar
n}=\nabla_{\bar m} \p_{\bar n} H_{\al \n}$. The second contribution
exists only in the presence of Yukawa coupling involving $A$, $B$ and
a gauge singlet $s$, and it is not described by the standard
supergravity Lagrangian. It involves a two-derivative coupling of two
anti-moduli to the auxiliary component of $\bar s$. The auxiliary field
propagates between this coupling and the Yukawa coupling to $A$ and
$B$, producing an effective interaction which contributes to ${\cal
A}(A^{\al},B^{\n},M^{\bar m},M^{\bar n})$. Such higher-derivative
couplings of auxiliary field and scalars are coming from a higher
weight F-term in superconformal supergravity.

Such higher derivative term can be expressed as\footnote{for
simplicity we will use the language of global supersymmetry although
they can be formulated in local supersymmetric way\cite{m_term}}
$$\int d^2\theta (\bar D^2f_1)(\bar D^2f_2)$$ where $\bar D^2$ is the
chiral projector and $f_1$ and $f_2$ are two arbitrary functions of
$\Phi$ and $\bar\Phi$. This would give rise to, among other terms,
coupling of the type $$\p\bar\phi^{\bar m} \p\bar\phi^{\bar n} h^{\bar
s} \underbrace{f_{\bar m\bar n}^{(1}
f_{\bar s}^{2)}}_{\equiv T_{\bar m\bar n\bar s}} $$ Such a term also
gives a contribution to the $\m$-term via the following diagram

\setlength{\unitlength}{1cm}
\begin{picture}(0,4.5)(0,-0.5)
\put(0,2.2){$<\chi^A\chi^B h^{\bar m}>\,\,\,\sim$} \put(5.2,2.2){\line(-
1,1){1}} \put(5.2,2.2){\line(-1,-1){1}} \put(5.2,2.2){\line(1,0){4}}
\put(3.7,3.1){$\chi^A$}\put(3.7,1.2){$\chi^B$}
\put(5.3,2.3){$s$}\put(6.6,2.3){$\Box \bar s$} \put(9.3,2.3){$h^{\bar
m}$}
\put(5.2,2.2){\circle*{0.1}}\put(7.2,2.2){\circle*{0.1}}
\put(6,3){${G^{s\bar s}\over \Box}$-propagator} \put(3,0.4){Yukawa
coupling $W_{ABs}$}
\put(6.8,1.1){$f_{\bar s}^{(1}f_{\bar m}^{2)}$} \end{picture}\\
\noindent Thus in the event of $h^{\bar m}$ acquiring a non-vanishing
v.e.v. one gets a mass term of the type $$\ti\m_{AB}=<h^{\bar
m}>_{v.e.v.} W_{ABs}G^{s\bar s}f_{\bar s}^{(1} f_{\bar m}^{2)}$$
which could be of phenomenological interest\cite{phenom} in the
study of low energy physics arising from string theory.

In \cite{m_term}, by using arguments similar to the ones used in
proving special geometry\footnote{i.e. using the N=2 superconformal
algebra of the bosonic sector\cite{geometry}}, it was shown that for
(2,2) models the function $T$ is related to the Yukawa coupling as \be
\nabla_{[\bar m}T_{\bar n]\bar k\bar s}
=\nabla_{[\bar m}\left( e^K \bar W_{\bar n]\bar k\bar s}\right)
\label{deriv}
\ee
where we have identified the charged indices ($\al,\n$, etc.) on the
right hand side in $W$, with the corresponding moduli indices $n,k$,
etc. (Recall that for (2,2) models for every (1,1) moduli there exist a
\boldmath $27$ \unboldmath and for (1,2) a \boldmath
$\overline{27}$ \unboldmath .) The right hand side of (\ref{deriv}) can
be computed by evaluating an amplitude $<\bar\chi^A\bar\chi^B
M^{\bar m}\bar s>$.

The purpose of this paper is to compute such amplitudes explicitly in
the case of certain orbifold model\footnote{in order to simplify
notation we evaluate the amplitudes of holomorphic states instead
antiholomorphic ones}. It turns out that in order to get It turns out
that in order to get
non-vanishing result all the 4 fields must be twisted fields. Recalling
that by SL(2,C) invariance we can fix 3 points on the sphere, this 4-
point correlation function still involves a 2-dimentional world sheet
integral over the position of one of the points. We will use the
following trick to evaluate this integral. If we take a holomorhic
derivative of the r.h.s. of (\ref{deriv}) with respect to an untwisted
modulus then the result is expected to be a total derivative in the
world sheet coordinate: $$\nabla_j e^{-K}\nabla_{\bar \imath}(e^K
W_{\bar A\bar B\bar s})= \nabla_j (\nabla_{\bar \imath}+\p_{\bar
\imath}K) W_{\bar A\bar B\bar s}$$ $$=[\nabla_j,\nabla_{\bar
\imath}] W_{\bar A\bar B\bar s}+ G_{j\bar \imath} W_{\bar A\bar
B\bar s}+ (\nabla_{\bar \imath}+\p_{\bar \imath}K)\nabla_j W_{\bar
A\bar B\bar s}$$ Since for our case $j$ and $\bar\imath$ are
untwisted and twisted indexes respectively, $G_{j\bar \imath}=0$. It
is also known that the holomorphic derivative of the Yukawa coupling
$\nabla_j W_{\bar A\bar B\bar s}$ is zero, so we have
$$
\nabla_j e^{-K}\nabla_{\bar \imath}(e^K W_{\bar A\bar B\bar s})=
R_{j\bar \imath\bar A A'}G^{A'\bar A'}W_{\bar A'\bar B\bar s}$$ \be
+R_{j\bar \imath\bar B B'}G^{B'\bar B'}W_{\bar A\bar B'\bar s}+
R_{j\bar \imath\bar s s'}G^{s'\bar s'}W_{\bar A\bar B\bar s'}
\label{2_deriv}
\ee
where the remaining terms correspond to factorized diagrams which
show us that the only contribution should come from the boundary of
the world sheet moduli space.
\setlength{\unitlength}{1cm}
\begin{picture}(15,5)(2.7,1.2)
\put(1.2,4){\line(-1,1){1}} \put(1.2,4){\line(-1,-1){1}}
\put(2.7,4){\line(1,1){1}} \put(2.7,4){\line(1,-1){1}}
\put(0.2,4){\line(1,0){2.5}}
\put(-0.2,4.8){$j$}\put(-0.2,3.9){$\bar \imath$}\put(-0.2,3.0){$\bar
A$} \put(3.8,4.8){$\bar B$}\put(3.8,3.0){$\bar s$}
\put(1.3,4.1){$A'$}\put(2.4,4.1){$\bar A'$}

\put(4.2,3.9){+}
\put(6.2,4){\line(-1,1){1}} \put(6.2,4){\line(-1,-1){1}}
\put(7.7,4){\line(1,1){1}} \put(7.7,4){\line(1,-1){1}}
\put(5.2,4){\line(1,0){2.5}}
\put(4.8,4.8){$j$}\put(4.8,3.9){$\bar \imath$}\put(4.8,3.0){$\bar B$}
\put(8.8,4.8){$\bar A$}\put(8.8,3.0){$\bar s$}
\put(6.3,4.1){$B'$}\put(7.4,4.1){$\bar B'$}

\put(9.2,3.9){+}
\put(11.2,4){\line(-1,1){1}} \put(11.2,4){\line(-1,-1){1}}
\put(12.7,4){\line(1,1){1}} \put(12.7,4){\line(1,-1){1}}
\put(10.2,4){\line(1,0){2.5}}
\put(9.8,4.8){$j$}\put(9.8,3.9){$\bar \imath$}\put(9.8,3.0){$\bar s$}
\put(13.8,4.8){$\bar A$}\put(13.8,3.0){$\bar B$}
\put(11.3,4.1){$s'$}\put(12.4,4.1){$\bar s'$} \end{picture}

Indeed in the following we will see an explicit example that
holomorphic derivative gives rise to a total derivative on the world
sheet as indicated by this general argument and this enables us to do
the integral explicitely.

The paper is organized as follows. In sect.2 we derive the 4-point
correlation function of the bosonic twist fields in the presence $g$,
$g^{-1}$, $h$ and $h^{-1}$ -twisted sectors using a technique
developed in \cite{conf_th_orb}. In sect.3 we give an example of the
non-zero correlation functions of the required type
$<\overline{27},1_s,27,1_m>$ and compute them as functions of
untwisted moduli (radii, etc.). In sect.4 we show that they form finite
dimensional representation of the duality group $SL_2(Z)$, acting on
the untwisted moduli space. The action of the duality group is shown
to be consistent with the action on the twisted fields characterized
by the fixed points set under the orbifold group. In sect.5 using the
above mentioned trick we find a differential equation for the
correlation function and discuss the uniqueness of its solution.
Conclusions are presented in sect.6. Calculation of some useful
integrals are given in the appendix.

\section{Correlation function of the bosonic twist fields}

We now present a calculation of a correlation function in which the
four twisted states at the points $z_{1}=0$, $z_{2}=x$, $z_{3}=1$ and
$z_{4}=\infty$ respectively are from $g,g^{-1},h$ and $h^{-1}$ -
twisted sectors. The correlation function to be calculated is \be
Z_{bos}(z_{i},\bar{z}_i) \equiv <\s_{g,f_{1}}(z_{1},\bar{z}_{1})
\s_{g^{-1},f_{2}}(z_{2},\bar{z}_{2})
\s_{h,f_{3}}(z_{3},\bar{z}_{3})
\s_{h^{-1},f_{4}}(z_{4},\bar{z}_{4})>
\label{cor_f}
\ee
The given twist field $\s_{g,f}$ determines how $X(z,\bar{z})$ is
rotated and translated when it is carried around that operator in the
complex $z$-plane:
\be
X(e^{2 \pi i}z,e^{-2 \pi i}\bar{z})=\theta^{j} X(z,\bar{z})+v
\label{mon_con}
\ee
where $v$ belongs to a coset of $\La$\footnote{by $\La$ we denoted
hexagonal lattice} which depends on the index $f$. If we split the field
$X$ into a classical piece $X_{cl}$ and a quantum fluctuation $X_{qu}$
then the correlation function divides into a sum over classical
solutions, times the quantum effective action, both evaluated in the
presence of twist fields:
\be
Z_{bos}=\sum_{<X_{cl}>} e^{-S_{cl}} Z_{qu} \label{tot_Z}
\ee
All global information needed to determine both the quantum Green
functions and the proper set of classical solutions can be determined
from the monodromy conditions for transporting $X_{qu}$ and $X_{cl}$
around collection of fields which have net twist zero.

\begin{figure}
\setlength{\unitlength}{1cm}
\begin{picture}(15,4)
\put(3,4){\circle*{0.1}}\put(8,4){\circle*{0.1}}
\put(3,1){\circle*{0.1}}\put(8,1){\circle*{0.1}}
\put(3,3.6){$z_1$}\put(8,3.6){$z_2$}
\put(3,0.6){$z_4$}\put(8,0.6){$z_3$}
\put(2,4){${\cal C}_1$}\put(9.5,2.5){${\cal C}_2$} \end{picture}
\caption{Two independent loops}
\label{2_loops}
\end{figure}
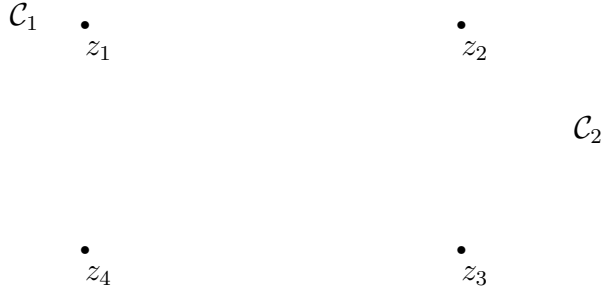

For our choice of closed loops shown in Fig.\ref{2_loops}, we have

\be
\oint_{{\cal C}_{a}} dX_{qu}=0,
\oint_{{\cal C}_{a}} dX_{cl}=v_{a},
a=1,2
\label{gl_con}
\ee
where
$$v_{1} \in (1-g)(f_{1}-f_{2}+\La_{1}-\La_{2})$$ \be
v_{2} \in (1-g)(1-h)(f_{3}-f_{2}+\La_{3}-\La_{2}) \label{v1_v2}
\ee
Taking the closed loop $\cal C$ to encircle all the vertex operators
gives the following constraints for the allowed fixed points:
\be
(1-g)(f_{1}-f_{2}+\La_{1}-\La_{2})
+(1-h)(f_{3}-f_{4}+\La_{3}-\La_{4})=0
\label{fix_points}
\ee

\subsection{Quantum part}
The first step in constructing the quantum piece of the correlation
function (\ref{cor_f}) is to find the Green functions in the presence of
the four twists
\be
g(z,w;z_i)\equiv {<-{1 \over 2} \p X(w) \p \bar X(z) \s_{g}(z_1)
\s_{g^{-1}}(z_2) \s_{h}(z_3) \s_{h^{-1}}(z_4)> \over <\s_{g}(z_1)
\s_{g^{-1}}(z_2) \s_{h}(z_3) \s_{h^{-1}}(z_4)>} \label{g_func}
\ee
\be
h(z,w;z_i)\equiv {<-{1 \over 2} \p X(w) \bar \p \bar X(\bar z)
\s_{g}(z_1) \s_{g^{-1}}(z_2) \s_{h}(z_3) \s_{h^{-1}}(z_4)> \over
<\s_{g}(z_1) \s_{g^{-1}}(z_2) \s_{h}(z_3) \s_{h^{-1}}(z_4)>}
\label{h_func}
\ee
Using asymptotic conditions as $z \rightarrow w$ and $z,w
\rightarrow z_i$ we find (we have used the following notation:
$g=e^{2\pi\al i}$, $h=e^{2\pi\beta i}$, $a=\al-1$, $b=\beta-1$, $a+c=-
1$ and $b+d=-1$) $$g(z,w;z_i)=\om_1(w)\om_2(z) \left\{
(b+1){(w-z_2)(w-z_4)(z-z_1)(z-z_3) \over (z-w)^2} \right.$$ $$ -
a{(w-z_1)(w-z_3)(z-z_2)(z-z_4) \over (z-w)^2} $$ $$\left. +(a-b){(w-
z_2)(w-z_3)(z-z_1)(z-z_4) \over (z-w)^2} +A(z_i,\bar z_i)\right\}$$
$$h(z,w;z_i)=\om_1(w) \bar \om_1(\bar z) B(z_i,\bar z_i)$$ where
$$\om_1(z)=(z-z_1)^a(z-z_2)^c(z-z_3)^b(z-z_4)^d$$ $$\om_2(z)=(z-
z_1)^c(z-z_2)^a(z-z_3)^d(z-z_4)^b$$ and $A(z_i,\bar z_i)$ and
$B(z_i,\bar z_i)$ will be determined later using the global monodromy
condition (\ref{gl_con}).

The operator product
$$ -{1 \over 2} \p_z X \p_w \bar X \sim {1 \over (z-
w)^2}+T(z)+\dots$$ under limit $w \rightarrow z$ gives us
$${<T(z)\s \dots \s > \over <\s \dots \s >}=\lim_{w \rightarrow z}
\left[ g(z,w)-{1 \over (z-w)^2} \right] $$ $$={1\over 2}\al (1-
\al)\left( {1 \over (z-z_1)}-{1 \over (z-z_2)}\right)^2 +{1\over
2}\beta (1-\beta)\left( {1 \over (z-z_3)}- {1 \over (z-
z_4)}\right)^2$$
$$+\beta (1-\al)\left( {1 \over (z-z_1)}-{1 \over (z-z_2)} \right)
\left( {1 \over (z-z_3)}-{1 \over (z-z_4)}\right)$$ $$+{A \over (z-
z_1)(z-z_2)(z-z_3)(z-z_4)}$$ Then the operator product
$$T(z)\s (z_2) \sim {h_{\s} \s (z_2) \over (z-z_2)^2} +{\p_{z_2} \s
(z_2) \over (z-z_2)}+ \dots$$ under limit $z \rightarrow z_2$ gives
rise to the differential equation $$\p_{z_2} \ln Z_{qu}(z_i,\bar z_i)=-
\al (1-\al){1 \over z_2-z_1} -\beta (1-\al) \left( {1 \over z_2-z_3} -
{1 \over z_2-z_4}\right) $$ \be
+{A \over (z_2-z_1)(z_2-z_3)(z_2-z_4)}
\label{z_2_eq}
\ee
Using now $SL_2(C)$ invariance we fix the location of the four vertex
operators: $z_1=0$, $z_2=x, z_3=1, z_4 \rightarrow \infty$. Then
(\ref{z_2_eq}) becomes
\be
\p_{x} \ln Z_{qu}(x,\bar x)=-\al (1-\al){1 \over x} -\beta (1-\al) {1
\over x-1}-{A \over x(1-x)} \label{dif_Z}
\ee
where
$$Z_{qu}(x,\bar x)=\lim_{z_{\infty} \rightarrow \infty}
|z_{\infty}|^{2\beta(1-\al)} <\s(0)\s(x)\s(1)\s(z_{\infty})>$$
$$A(x,\bar x)=\lim_{z_{\infty} \rightarrow \infty}(-z_{\infty}^{-1})
A(0,x,1,z_{\infty})$$
The global monodromy condition (\ref{gl_con}) gives equation to
determine $A$ and $B$:
$$\oint_{{\cal C}_a}dz g(z,w)+\oint_{{\cal C}_a}d\bar z h(\bar
z,w)=0$$ which became after fixing $z_i$ and dividing by $\om_1(w)$:
$$A\oint_{{\cal C}_a}dz \om_2(z)+
B\oint_{{\cal C}_a}d\bar z \bar \om_1(\bar z) =\oint_{{\cal C}_a}dz
\om_2(z)$$
$$\times \left[ -(b+1){(w-x)z(z-1)\over (z-w)^2} +a {w(w-x)(z-
x)\over (z-w)^2} +(b-a){(w-x)(w-1)z\over (z-w)^2} \right] $$ Choosing
$w=1$ we can find that
$$A\oint_{{\cal C}_1}dz \om_2(z)+
B\oint_{{\cal C}_1}d\bar z \bar \om_1(\bar z)= x(1-x){d\over dx}
\oint_{{\cal C}_1}dz \om_2(z)$$ and for $w=0$
$$A\oint_{{\cal C}_2}dz \om_2(z)+
B\oint_{{\cal C}_2}d\bar z \bar \om_1(\bar z)= x(1-x){d\over dx}
\oint_{{\cal C}_2}dz \om_2(z)$$ Solving this equations we find
\be
A=x(1-x)\p_x \ln I(x,\bar x)
\label{sol_A}
\ee
\be
B={x(1-x)\bar\g \left( \oint_{{\cal C}_1}\om_2 \right)^2 \over
I(x,\bar x)} {d\tau\over dx}
\label{sol_B}
\ee
where
\be
I(x,\bar x)=\oint_{{\cal C}_1}\omega_2\oint_{{\cal
C}_2}\bar\omega_1 -\oint_{{\cal C}_2}\omega_2\oint_{{\cal
C}_1}\bar\omega_1 =\oint_{{\cal C}_1}\omega_2\oint_{{\cal
C}_1}\bar\omega_1 \,\bar \g\, (\bar\tau-\tau )
\label{def_I}
\ee
Substituting (\ref{sol_A}) into (\ref{dif_Z}) and integrating it we get
$$Z_{qu}(x,\bar x)=const {|x|^{-2\al (1-\al )}|1-x|^{-2\beta(1-\al)}
\over \left| \oint_{{\cal C}_1}\omega_2 \right|^2 \tau_2}$$ \be
=const {|x|^{-2\al (1-\al )}|1-x|^{-2\al(1-\beta)} \over \left| F(1-
\beta,\al;1;x) \right|^2 \tau_2} \label{bos_qu}
\ee
where we have used the $x\leftrightarrow\bar x$ symmetry of
$Z_{qu}(x,\bar x)$ to fix the $\bar x$-dependence of the integration
constant.
 From (\ref{sol_B}) we find
\be
h(z,w|z_i)={i\over 2 \tau_2}x(1-x)\om_1(w)\bar\om_1(\bar z)
{\oint_{{\cal C}_1}\omega_2\over \oint_{{\cal C}_1}\bar\omega_1}
{d\tau\over dx}
\label{gr_f_h}
\ee

\subsection{Classical part}
The $z$ and $\bar z$ derivatives of the classical solutions can be
written as
$$ \p X_{cl}(z)=a\om_1(z),\,\bar \p X_{cl}(\bar z)=\bar b \bar
\om_2(\bar z)$$ $$ \p \bar X_{cl}(z)=b\om_2(z),\,\bar \p \bar
X_{cl}(\bar z)= \bar a \bar \om_1(\bar z)$$
Substituting this into (\ref{gl_con}) and solving it we find: $$a={v_1
\oint_{{\cal C}_2}\bar \om_2-v_2 \oint_{{\cal C}_1}\bar \om_2
\over \oint_{{\cal C}_1}\om_1 \oint_{{\cal C}_2}\bar \om_2-
\oint_{{\cal C}_2}\om_1 \oint_{{\cal C}_1}\bar \om_2}$$ $$\bar
b={v_1 \oint_{{\cal C}_2} \om_1-v_2 \oint_{{\cal C}_1} \om_1 \over
\oint_{{\cal C}_1}\bar \om_2 \oint_{{\cal C}_2} \om_1- \oint_{{\cal
C}_2}\bar \om_2 \oint_{{\cal C}_1} \om_1}$$ Normalizing $\om_1$
and $\om_2$ and using definition of $\tau$ given in Appendix we can
rewrite
$$a={v_1 (\bar \tau-\al)\g -v_2 \over \g(\bar \tau-\tau)}$$ $$\bar
b={v_1 (\tau-\al)\g -v_2 \over \g(\tau-\bar \tau)}$$ The classical
action is given by
$$S_{cl}={1 \over 4\pi}\int_C d^2 z \left[ R^2 \right. \left( \p X_{cl}
\bar \p \bar X_{cl}+\bar \p X_{cl} \p \bar X_{cl}\right)$$ $$+ib \left(
\p X_{cl} \bar \p \bar X_{cl}-\bar \p X_{cl} \p \bar X_{cl} \right)
\left. \right] $$
$$={-i\over 2\pi\sqrt{3}}\left[ T\int d^2z\p X_{cl} \bar\p\bar X_{cl} -
\bar T\int d^2z \bar\p X_{cl} \p\bar X_{cl}\right]$$ where
$$T={\sqrt{3}\over 2}(i R^2-b)$$
Using results of Appendix after some algebra, the action for the
classical solutions becomes
\be
S_{cl}={1 \over 8\pi |\g|^2} \left[ R^2 \right. \left(
|v_1\g|^2\tau_2+{1\over \tau_2} |v_1\g(\tau_1 -\al)+v_2|^2 \right) $$
$$-b \left( |v_1\g|^2 (\bar \al-\al)-v_1\g \bar v_2+\bar v_1 \bar \g
v_2 \right) \left. \right]
\label{cl_act}
\ee
where
$$\tau(x)\equiv \tau_1+i\tau_2
={\oint_{{\cal C}_2} \om_1 \over \g \oint_{{\cal C}_1} \om_1}-\al$$
\be
={i \over 2 \sqrt{3}}\left( {\sqrt{3}\over \pi}(1-x)^{1/6}
{\Gamma({1\over 2})\Gamma({2\over 3})\over \Gamma({7\over 6})}
{F({2\over 3},{1\over 2};{7\over 6};1-x) \over F({1\over 2},{1\over
3};1;x)} +{1\over 2}\right)
\label{tau}
\ee
The classical part of the correlation function is given by
$$Z_{cl}=\sum_{v_1,v_2}e^{-S_{cl}}$$
where sum is over cosets $v_1$ and $v_2$ determined in
(\ref{v1_v2}). Substituting $v_1=\g (\ep_1+\La)$ and $v_2=\g
(\ep_2+\La)$ we found that $\ep_2=f_{23}$\footnote{we used
notation $f_{12}=f_1-f_2$,etc.} and $\ep_1$ satisfies the constraint
\be
\ep_1+\La={1\over 1-h}(f_{12}+\La)\bigcap ({1\over 1-g}(f_{34}+\La)
\label{eps}
\ee
After Poisson resummation on $v_2$ we get \be
Z_{cl}={8\pi|\g|^2\over R^2}\tau_2\sum_{v\in \Lambda,p\in
\Lambda^*} e^{2\pi i (\ep_2-\g\al v-\g\al\ep_1)p}
w^Q_L \bar w^Q_R
\label{bos_cl}
\ee
where $w=e^{i\pi\tau},\bar w=e^{-i\pi\bar\tau}$ and $$Q_L=|{p\over
R}+{(v+\ep_1)\g\over 2R}(R^2-ib)|^2$$ $$Q_R=|{p\over R}-
{(v+\ep_1)\g\over 2R}(R^2+ib)|^2$$ \section{The correlation function
$<\overline{27},1_s,27,1_m>$}

We start from $E_8\times E_8$ heterotic string compactified on the
6-dimensional orbifold, which is a direct sum of three two-
dimensional orbifolds\footnote{for descripsion of orbifolds and
notation see \cite{conf_th_orb}}:
$$Z_6=Z_2^{(6)}\times Z_2^{(3)}\times Z_2^{(2)}$$ where the lattices
for the first two are hexagonal (we denote it by $\Lambda$) and for
the third is arbitrary. Point groups are generated by rotations
$\theta$ of order 6, 3 and 2 respectively.

To simplify calculations is useful to bosonize all fermions. Using
fermionic charge conservation, we found the states $\overline{27}$,
$27$, $1_m$ and $1_s$, whose correlation function of the type
$<f,f,b,b>$ is not zero. The states $\overline{27}$ and $1_s$ come
from $g^2$, $g^{-2}$ twisted sectors respectively and we choose them
to be in the Ramond sector. States $27$ and $1_m$ come from $g^3$
twisted sector and they are in the Neveu-Schwarz sector.

\begin{figure}
\setlength{\unitlength}{1cm}
\begin{picture}(15,4)
\put(2,4){\line(1,0){4}}\put(2,4){\line(3,-5){2}} \put(6,4){\line(-3,-
5){2}}\put(4,0.6667){\line(1,0){4}} \put(6,4){\line(3,-5){2}}
\put(4,0.6667){\circle*{0.1}}\put(4,2.8889){\circle*{0.1}}
\put(6,1.7778){\circle*{0.1}}
\put(4,0.2){0}\put(4,3.1){$\kappa_2$}\put(6,2){$\kappa_1$}
\put(4,0.6667){\circle*{0.1}}\put(6,0.6667){\circle*{0.1}}
\put(3,2.3333){\circle*{0.1}}\put(5,2.3333){\circle*{0.1}}
\put(4,0.2){0}\put(6,0.2){$\rho_1$}
\put(2.6,2.1){$\rho_3$}\put(5.1,2.1){$\rho_2$} \end{picture}
\caption{The fixed points for the ${2\pi\over 3}$ and ${2\pi\over 4}$
rotations on the hexagonal lattice $\Lambda$} \label{lattice}
\end{figure}
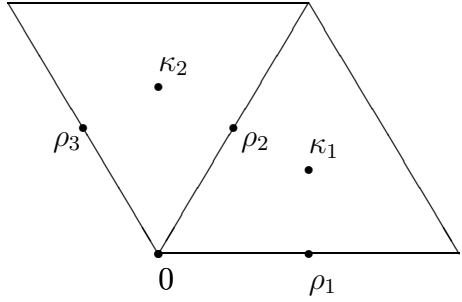

Vertex operators for states are
\begin{eqnarray*}
\overline{27}:V_{-1/2}(z_1)&=&e^{-\phi/2}e^{i \al H}e^{ip \ti H} \s^1
\s^2 \s^3 (z_1) |B> \\
\al&=&(-{1 \over 6},{1 \over 6},-{1 \over 2},{1 \over 2},-{1 \over
2})\\ p&=&({2 \over 3},{1 \over 3},0|1,0^4)
\end{eqnarray*}
\begin{eqnarray*}
1_s:V_{-1/2}(z_2)&=&e^{-\phi/2}e^{i \al H}e^{ip \ti H} \s^1 \s^2 \s^3
(z_2) |B> \\
\al&=&({1 \over 6},-{1 \over 6},-{1 \over 2},-{1 \over 2},{1 \over
2})\\ p&=&(-{2 \over 3},-{1 \over 3},1|0^5)
\end{eqnarray*}
\begin{eqnarray*}
27:V_{-1}(z_3)&=&e^{-\phi}e^{i \al H}e^{ip \ti H}\s^1 \s^2 \s^3 (z_3)\\
\left( |0> or |C> \right) \\
\al&=&({1 \over 2},0,{1 \over 2},0,0)\\
p&=&(-{1 \over 2},0,-{1 \over 2}|-1,0^4) \end{eqnarray*}
\begin{eqnarray*}
1_m:V_{0}(z_4)&=&{\lim}_{w\rightarrow z_4}\oint_{C_{z_4}} d\bar z
(e^{\phi}e^{-iH_1}\p X^1)(w) (e^{i \ti H_1} \bar \p \bar X^1)(\bar z)\\ &
& (e^{-\phi}e^{i\al H}e^{ip\ti H}\s^1\s^2\s^3)(z_4) \left( |0> or |C>
\right) \\
\al&=&({1 \over 2},0,{1 \over 2},0,0)\\
p&=&(-{1 \over 2},0,-{1 \over 2}|0^5)
\end{eqnarray*}
In the expression for the state $1_m$ we leave only non-vanishing
component of $T_F$. The ground states $|B>$ and $|C>$ are eigenstates
of the group action $g$ and they are linear combination of the fixed
points: $$|B>=|\kappa_1>-|\kappa_2>$$
$$|C>=|\rho_1>+|\rho_2>+|\rho_3>$$
where $\kappa_1$, $\kappa_2$ are the fixed points for the $2\pi /3$
rotation and $\rho_1$, $\rho_2$, $\rho_3$ are the fixed points for the
$\pi$ rotation on the hexagonal lattice $\La$ (see Fig.\ref{lattice}).
The correlation functions with different ground states form the
multiplet $$\left(\begin{array}{c}
<B,B,O,O> \\ <B,B,C,O> \\ <B,B,O,C> \\ <B,B,C,C> \end{array}\right)$$
under the modular group transformations, which we will show later.

The correlation function $<\overline{27},1_s,27,1_m>$ is $$Z(x,\bar
x)=<V_{-1/2}(0)V_{-1/2}(x)V_{-1}(1)V_0(z_{\infty})>$$
$$=\lim_{w\rightarrow z_4} \oint_{C_{z_4}}d\bar z <e^{-
\phi/2}(0)e^{-\phi/2}(x)e^{-\phi}(1)e^{-\phi}(z_{\infty})e^{\phi}(w)>$$
$$\times <e^{i\s}(0) e^{i\s}(1) e^{i\s}(z_{\infty})\times c.c.>$$
$$\times <e^{i\al H}(0)e^{i\al H}(x)e^{i\al H}(1)e^{i\al
H}(z_{\infty})e^{-iH_1}(w)>$$ $$\times <e^{ip \ti H}(0)e^{ip \ti H}(\bar
x)e^{ip \ti H}(1)e^{ip \ti H}(z_{\infty}) e^{i\ti H_1}(\bar z)>$$
$$\times <\s_{1/3}^1(0)\s_{-1/3}^1(x,\bar
x)\s_{1/2}^1(1)\s_{1/2}^1(z_{\infty}) \p X^1(w)\bar\p\bar X^1(\bar
z)>$$
\be
\times <\s_{2/3}^2(0)\s_{-2/3}^2(x,\bar x)>
<\s_{1/2}^3(1)\s_{1/2}^3(z_{\infty})> \label{full_Z}
\ee
Here we had set space-time momenta to zero.

Correlations of all fields except the bosonic twists are found from
the formula:
$$<e^{ik_1X}(z_1)\dots e^{ik_nX}(z_n)=
\prod_{1\leq i<j\leq n} (z_i-z_j)^{k_i k_j}$$

The correlation of the bosonic twist fields is splitted into a sum of
classical and quantum parts.
$$<\p X(w)\bar\p\bar X(\bar z) \s\dots\s>= <\p X_{cl}(w)\bar\p\bar
X_{cl}(\bar z) \s\dots\s>$$ \be
+<\p X_{qu}(w)\bar\p\bar X_{qu}(\bar z) \s\dots\s> \label{bos_cor}
\ee
In order to determine the classical part we will use the following
trick. Let's take derivative of $Z_{bos}$ with respect to $T$:
$${\p\over \p T}Z_{bos}=-{1\over 4\pi\sqrt{3}}\sum_{<X_{cl}>} \int
d^2 z \p X_{cl}(z)\wedge\bar\p\bar X_{cl}(\bar z)Z_{qu}e^{-S_{cl}}$$
\be
={i\tau_2\over 2\pi\sqrt{3}}
\oint_{{\cal C}_1}\om_1\oint_{{\cal C}_1}\bar\om_1 \sum_{<X_{cl}>}
|a|^2 Z_{qu}e^{-S_{cl}}
\label{der_Z}
\ee
Then we see that
$$<\p X(w)_{cl}\bar\p\bar X_{cl}(\bar z)\s\dots\s>=
\omega_1(w)\bar\omega_1(\bar z)\sum |a|^2 Z_{qu}e^{-S_{cl}}$$ \be
={2\pi\sqrt{3}\over i\tau_2}\,{\omega_1(w)\bar\omega_1(\bar z)
\over \oint_{{\cal C}_1}\om_1\oint_{{\cal C}_1}\bar\om_1} {\p\over
\p T}Z_{bos}
\label{Z_cl}
\ee
The quantum part $<\p X_{qu}(w)\bar\p\bar X_{qu}(\bar z) \s\dots\s>$
is found from (\ref{gr_f_h}) and (\ref{h_func}) $$<\p
X_{qu}(w)\bar\p\bar X_{qu}(\bar z) \s\dots\s>= -2h(\bar
z,w)Z_{qu}Z_{cl}$$
\be
= x(1-x){\omega_1(w)\bar\omega_1(\bar z) \over
i\tau_2,\oint_{{\cal C}_1}\om_1\oint_{{\cal C}_1}\bar\om_1} {\p
\tau\over \p x} \oint_{{\cal C}_1}\om_2\oint_{{\cal C}_1}\om_1
Z_{bos} \label{Z_qu}
\ee
Substituting (\ref{Z_cl}) and (\ref{Z_qu}) in (\ref{bos_cor}) we have
$$<\p X(w)\bar\p\bar X(\bar z)\s\dots\s>=
{\omega_1(w)\bar\omega_1(\bar z) \over
\tau_2\, \left| \oint_{{\cal C}_1} \om_1 \right|^2} {1\over R^2}
{\p\over \p T} \left( R^2 Z_{bos}\right)$$ $$=const \,
{\omega_1(w)\bar\omega_1(\bar z) \over \tau_2 \, \left|
F(1/2,1/3,1,x) \right|^2} $$ \be
\times {1\over R^2}{\p\over \p T}
\left[ {|x|^{-4/9}|1-x|^{-1/3}\over |F(1/2,1/3,1,x)|^2}
\sum_{v\in\Lambda,p\in\Lambda^*}
e^{2\pi i (\ep_2-\g\al v-\g\al\ep_1)p} w^{Q_L} \bar w^{Q_R}\right]
\label{X_X_Z}
\ee

The full correlation function (\ref{full_Z}) is then integrated over the
complex plane and gives the scattering apmlitude for four massless
twisted states:
$$A_{\ep_j}(T,\bar T)=\int d^2 x Z(x,\bar x) =const\,\int d^2 x {1\over
\tau_2}\,{x^{-1}(1-x)^{-5/6}\bar x^{-1}(1-\bar x)^{-1/3} \over
|F(1/2,1/3;1;x)|^4}$$
$$
\times {1\over R^2} {\p\over \p T}
\sum_{v\in\Lambda,p\in\Lambda^*} e^{2\pi i (\ep_2-\g\al v-
\g\al\ep_1)p} w^{Q_L} \bar w^{Q_R} $$
Recalling that
$${d\tau\over dx}={1\over 2\pi i}{x^{-1}(1-x)^{-5/6} \over
F(1/2,1/3;1;x)^2}$$
we rewrite
\be
A_{\ep_j}(T,\bar T)={1\over R^2}\int d^2 x {1\over
\tau_2}\,{d\tau\over dx}{d\bar\tau\over d\bar x}(1-\bar
x)^{1/2}{\p\over \p T} C(\ep_1,\ep_2) \label{ampl}
\ee
where
\be
C(\ep_1,\ep_2)=\sum_{v\in \Lambda,p\in \Lambda^*} e^{2\pi i
(\ep_2-\g\al v-\g\al\ep_1)p}
w^{Q_L} \bar w^{Q_R}
\label{def_C}
\ee

\section{Modular transformation}

If we decompose $v=v_0+\rho$ and $p=p_0+c+\m$, where $v_0\in
2\Lambda, \rho\in\Lambda/2\Lambda$, $p_0\in 2\g\Lambda^*,\m\in
\g\Lambda^*/2\g\Lambda^*, c\in \Lambda^*/\g\Lambda^*$ then
(\ref{def_C}) becomes $$C(\ep_1,\ep_2)=\sum_{\rho,c,\m}
\sum_{v_0,p_0} e^{2\pi i(\ep_2-\g\al (v_0+\rho)-
\g\al\ep_1)(p_0+c+\m)}w^{Q_L} \bar w^{Q_R}$$ \be
=\sum_{\rho,c,\m} e^{2\pi i(\ep_2-\al\g\ep_1)c} e^{-2\pi i\al\g(\rho
c+\ep_1)\m)}Z(\ep_1+\rho,c+\m) \label{ext_C}
\ee
where
$$Z(\ep_1+\rho,c+\m)=\sum_{v_0,p_0}w^{Q_L} \bar w^{Q_R}$$
FFrom (\ref{ext_C}) we find it's Fourier transformation
\be
\sum_{\ep_2}C(\ep_1,\ep_2) e^{-2\pi i(\ep_2-\al\g\ep_1)\ti
c}=12\sum_{\rho, \ti\m}e^{-2\pi i\al\g(\rho\ti
c+\ep_1)\ti\m)}Z(\ep_1+\rho,\ti c+\ti\m) \label{fur_C}
\ee
The correlation function should be invariant under modular
transformations $$T\rightarrow {aT+b\over cT+d}$$
where $T={\sqrt{3}\over2}(iR^2-b)$ and $a,b,c,d$ form an element of
$SL_2(Z)$.

Modular transformations can be generated by two basic
transformations: $${\cal S}:T\rightarrow -{1\over T}\Leftrightarrow
\left\{ \begin{array}{l}
R^2+ib\rightarrow {4\over 3(R^2+ib)}\\
R^2-ib\rightarrow {4\over 3(R^2-ib)}
\end{array} \right.
\Rightarrow R\rightarrow {2R\over\sqrt{3}(R^4+b^2)^{1/2}}$$ and
$${\cal T}:T\rightarrow T+1 \Leftrightarrow \left\{ \begin{array}{l}
b\rightarrow b-{2\over\sqrt{3}}\\
R\rightarrow R
\end{array} \right. $$

\subsection{{\cal S}-transformation}
We find that under {\cal S}-transformation
$$Z(\ep_1+\rho,c+\m)\stackrel{{\cal S}}{\rightarrow}
Z(\ti\ep_1+\ti\rho,\ti c+\ti\m)$$
where
$$\ti v_0={\theta\over 2}p_0,\ti\ep_1={\theta\over 2}c,
\ti\rho={\theta\over 2}\m$$
$$\ti p_0=2v_0,\ti c=2 \ep_1,\ti \m=2\rho$$ After some calculations
we find
$$C(\ep_1,\ep_2)\stackrel{{\cal S}}{\rightarrow}{1\over12}
\sum_{\ti \ep_1,\ti \ep_2} e^{4\pi i(\theta^{-1}\ti\ep_1\ep_2-
\ep_1\ti\ep_2-\ep_1\theta\ti\ep_1)} C(\ti\ep_1,\ti\ep_2)$$
Knowing that the solutions of (\ref{eps}) for the case $g=\theta^2$,
$h=\theta^3$ ($\theta=e^{2\pi i/6}$) is $$\ep_2=f_{23},
\ep_1=\theta^2 f_{34}-f_{12}$$ and substituting them back we get
$$
<f_1,f_2,f_3,f_4>\stackrel{{\cal S}}{\rightarrow} {1\over12}\sum
e^{4\pi i(f_1\ti f_1-f_2\ti f_2-f_3\theta\ti f_3 -f_4\theta\ti
f_4)}<\ti f_1,\ti f_2,\ti f_3,\ti f_4> $$
from which we can deduce that the modular transformation acts on
states separately:
$$<f_1>\stackrel{{\cal S}}{\rightarrow} {1\over \sqrt{3}} e^{4\pi
if_1\ti f_1}<\ti f_1>$$
$$<f_2>\stackrel{{\cal S}}{\rightarrow} {1\over \sqrt{3}} e^{-4\pi
if_2\ti f_2}<\ti f_2>$$
$$<f_3>\stackrel{{\cal S}}{\rightarrow} {1\over 2} e^{-4\pi
if_3\theta\ti f_3}<\ti f_3>$$
$$<f_4>\stackrel{{\cal S}}{\rightarrow} {1\over 2} e^{-4\pi
if_4\theta\ti f_4}<\ti f_4>$$
For the physical states the transformation become

$<B>\stackrel{{\cal S}}{\rightarrow} -i<B>$ -for $g^2$ twisted sector

$<B>\stackrel{{\cal S}}{\rightarrow} i <B>$ -for $g^{-2}$ twisted
sector

\noindent and

$\left(\begin{array}{c}
<O> \\ {1\over\sqrt{3}}<C>
\end{array}\right) \stackrel{{\cal S}}{\rightarrow}{1\over 2}
\left(\begin{array}{cc}
1	& \sqrt{3}\\
\sqrt{3} & -1	\\
\end{array}\right)
\left(\begin{array}{c}
<O> \\ {1\over\sqrt{3}}<C>
\end{array}\right)$ -for $g^{3}$ twisted sector

\noindent Finally we get that the multiplet of the correlation
functions transforms
$$\left(\begin{array}{c}
<B,B,O,O> \\ <B,B,C,O> \\ <B,B,O,C> \\ <B,B,C,C> \end{array}\right)
\stackrel{{\cal S}}{\rightarrow}{1\over 4} \left(\begin{array}{cccc}
1& 1& 1& 1\\
3&-1& 3&-1\\
3& 3&-1&-1\\
9&-3&-3& 1
\end{array}\right)
\left(\begin{array}{c}
<B,B,O,O> \\ <B,B,C,O> \\ <B,B,O,C> \\ <B,B,C,C> \end{array}\right)$$
Defining the quantity $G(T,\bar T)=R^2 A_{\ep_j}(T,\bar T)$ we find
the following transformation
\be
G\left(\begin{array}{c}
<O,O> \\ <C,O> \\ <O,C> \\ <C,C>
\end{array}\right) \stackrel{{\cal S}}{\rightarrow}{T^2\over 4}
\left(\begin{array}{cccc}
1& 1& 1& 1\\
3&-1& 3&-1\\
3& 3&-1&-1\\
9&-3&-3& 1
\end{array}\right)
G\left(\begin{array}{c}
<O,O> \\ <C,O> \\ <O,C> \\ <C,C>
\end{array}\right)
\label{S-transf}
\ee

\subsection{{\cal T}-transformation}
Under {\cal T}-transformation
$$Z(\ep_1+\rho,c+\m)\stackrel{\cal T}{\rightarrow} Z(\ep_1+\rho,\ti
c+\ti\m)$$ where
$$\ti p_0=p_0+2\theta v_0,\ti
c=c+2\theta\ep_1,\ti\m=\m+2\theta\rho$$ After some calculations
we find
$$C(\ep_1,\ep_2)\stackrel{\cal T}{\rightarrow} e^{-2\pi
i(2\ep_2+\theta\ep_1)\theta\ep_1}C(\ep_1,\ep_2)$$ and
\be
<f_1,f_2,f_3,f_4>\stackrel{\cal T}{\rightarrow} e^{2\pi i(-f_1
f_1+f_2 f_2+f_3 f_3+f_4 f_4)} <f_1,f_2,f_3,f_4> \label{T_tran}
\ee
 From (\ref{T_tran}) we can deduce, as in the previous case, that
the modular transformation acts on states separately:
$$<f_1>\stackrel{{\cal T}}{\rightarrow} e^{-2\pi if_1 f_1}<f_1>$$
$$<f_2>\stackrel{{\cal T}}{\rightarrow} e^{2\pi if_2 f_2}<f_2>$$
$$<f_3>\stackrel{{\cal T}}{\rightarrow} e^{2\pi if_3 f_3}<f_3>$$
$$<f_4>\stackrel{{\cal T}}{\rightarrow} e^{2\pi if_4 f_4}<f_4>$$
which for physical states become

$<B>\stackrel{{\cal T}}{\rightarrow} e^{2\pi i/3}<B>$ -for $g^2$
twisted sector

$<B>\stackrel{{\cal T}}{\rightarrow} e^{-2\pi i/3}<B>$ -for $g^{-2}$
twisted sector

\noindent and

$\left(\begin{array}{c}
<O> \\ <C>
\end{array}\right) \stackrel{{\cal T}}{\rightarrow}
\left(\begin{array}{cc}
1 & 0 \\
0 & -1 \\
\end{array}\right)
\left(\begin{array}{c}
<O> \\ <C>
\end{array}\right)$
-for $g^{3}$ twisted sector

\noindent The multiplet of the correlation functions transforms as
follows
$$\left(\begin{array}{c}
<B,B,O,O> \\ <B,B,C,O> \\ <B,B,O,C> \\ <B,B,C,C> \end{array}\right)
\stackrel{\cal T}{\rightarrow} \left(\begin{array}{cccc}
1& 0& 0& 0\\
0&-1& 0& 0\\
0& 0&-1& 0\\
0& 0& 0& 1
\end{array}\right)
\left(\begin{array}{c}
<B,B,O,O> \\ <B,B,C,O> \\ <B,B,O,C> \\ <B,B,C,C> \end{array}\right)$$
Under ${\cal T}$-transformation $G(T,\bar T)$ transforms as \be
G\left(\begin{array}{c}
<O,O> \\ <C,O> \\ <O,C> \\ <C,C>
\end{array}\right) \stackrel{\cal T}{\rightarrow}
\left(\begin{array}{cccc}
1& 0& 0& 0\\
0&-1& 0& 0\\
0& 0&-1& 0\\
0& 0& 0& 1
\end{array}\right)
G\left(\begin{array}{c}
<O,O> \\ <C,O> \\ <O,C> \\ <C,C>
\end{array}\right)
\label{T-transf}
\ee
Since ${\cal S}$ and ${\cal T}$ transformations generate all modular
group, from (\ref{S-transf}) and (\ref{T-transf}) we deduce that the
correlation functions (\ref{ampl}) form finite dimensional
representation of the duality group $SL_2(Z)$.
\section{Integration of the correlation function}

In order to integrate the amplitude (\ref{ampl}) we take a derivative
of $G(T,\bar T)$ with respect to $\bar T$ and refering to the
following formula\cite{2_deriv}
$${\p^2\over \p \bar T\p T}(w^{Q_L}\bar w^{Q_R}) =-{4\tau_2\over (T-
\bar T)^2}\p_{\tau}\p_{\bar\tau}(w^{Q_L}\bar w^{Q_R}\tau_2)$$ we
find
$${\p\over \p\bar T}G(T,\bar T)=\int d^2x {1\over \tau_2}{d\tau\over
dx} {d\bar\tau\over d\bar x}(1-\bar x)^{1/2}{\p^2\over \p\bar T\p
T}C(x,\bar x)$$ $$=-{4\over(T-\bar T)^2}\int d^2x \p_x[
{d\bar\tau\over d\bar x}(1-\bar x)^{1/2} \p_{\bar\tau}(C(x,\bar
x)\tau_2)]$$
The later integral can be written as a surface integral, turning out to
be a sum of the surface integrals at $0,1,\infty$. $${\p\over \p\bar
T}G(T,\bar T)=-{2\sqrt{2}\pi\over(T-\bar T)^2} \sum_{x_0\in \{
0,1,\infty\} } (\bar x-\bar x_0)^{1/2} {d\bar\tau\over d\bar x}(1-\bar
x)^{1/2} \p_{\bar\tau}(C(x,\bar x)\tau_2)]_{x=x_0,\bar x=\bar x_0}$$

The first term of the above sum at $x_0=1$ does not give any
contribution while the terms at $x_0=0$ and $x_0=\infty$ give rise to
a non-zero result. Recalling $\tau\rightarrow {i\over 2\pi}\log
\left({12\over x}\right)$ then $x\rightarrow 0$ we find that the
second term of the sum at $x_0=0$ equal
$$-{2\sqrt{2}\pi i\over(T-\bar T)^2}$$
where the only non-zero contribution is from sector $Q_L=Q_R=0$.

To find the contribution at $x_0=\infty$ the integral over $x$-plane is
written as an integral over $\tau$-plane: $$-{4\over(T-\bar T)^2}\int
d^2\tau {1\over \tau_2}(1-\bar x(\bar \tau))^{1/2}
\p_{\tau}\p_{\bar\tau}(C(\tau,\bar \tau)\tau_2)]$$ $$=-{2\sqrt{2}\pi
\over(T-\bar T)^2}(\bar\tau-\bar\tau_0) (1-\bar x(\bar \tau))^{1/2}
\p_{\bar\tau}(C(\tau,\bar \tau)\tau_2)]_{\tau=\tau_0}$$ where
$$\tau_0=-{1\over 2}+{i\over 4\sqrt{3}}$$ Expanding the functions
$(1-\bar x(\bar \tau))^{1/2}$ and $\p_{\bar\tau}(C(\tau,\bar
\tau)\tau_2)$ in powers of $(\bar\tau-\bar\tau_0)$ we find that the
third term of the sum at $x_0=\infty$ is
$$-{\pi\over(T-\bar T)^2}{b^3\sqrt{2}\over 12\sqrt{3}}
\sum_{v\in\Lambda,p\in\Lambda^*}
e^{2\pi i (\ep_2-\g\al v-\g\al\ep_1)p} w^{Q_L} \bar w^{Q_R}P(Q_R)$$
where
$$P(Q_R)={\pi^3\over 8\sqrt{3}}Q_R^3-{9\pi^2\over 4}Q_R^2+
9\sqrt{3}\pi Q_R-18$$

The first term vanishes according to the general argument mentioned
in above. The corresponding term is $R_{\bar T m \overline{27}\,
\overline{27}'^{*}}G_{\overline{27}'^{*}\,\overline{27}'}
W_{\overline{27}'\,27\,s}$ and it happens in our model that there is
not a such Yukawa coupling. The reason why the second term of the
sum does not involve a lattice sum is that the corresponding term is
$R_{\bar T m 27\,27'^{*}}G_{27'^{*}\,27'} W_{27'\,\overline{27}\,s}$
with the untwisted intermediat state $27'$. For the third term the
corresponding term is $R_{\bar T m\, s\,s'^{*}}G_{s'^{*}\,s'}
W_{27\,\overline{27}\,s'}$ so the intermediat state $s'$ comes from
the twisted sector which gives the lattice sum.

As a final result we have
$${\p\over \p\bar T}G(T,\bar T)=
-{\sqrt{2}\pi i\over(T-\bar T)^2}
-{\pi\over(T-\bar T)^2}{b^3\sqrt{2}\over 12\sqrt{3}}$$ \be
\times
\sum_{v\in\Lambda,p\in\Lambda^*}
e^{2\pi i (\ep_2-\g\al v-\g\al\ep_1)p} w^{Q_L} \bar w^{Q_R}P(Q_R)
\label{dif_eq}
\ee

Solving this differetial equation one can in principle find a solution up
to a holomorphic functions of $T$. These functions should transforms
under the modular group according (\ref{S-transf}) and (\ref{T-
transf}). This transformation law implies that the additional terms
are modular forms of weight $k=2$ of the subgroup $\Gamma(2)$ of
the full modular group. According to \cite{Rankin} they are only of the
form $\al S+\beta V$, where $S=\theta_3^4+\theta_4^4$,
$V=\theta_2^4$ and $\theta$'s are theta functions. Using the
transformation properties

${\cal T}$:$S \rightarrow S$, $V \rightarrow -V$

${\cal S}$:$S \rightarrow -{1\over 2}(S+3V)$, $V \rightarrow -
{1\over 2}(S-V)$

\noindent and requiring the holomorphic ambiguities to satisfy the
trasformations (\ref{S-transf}) and (\ref{T-transf}) we found that
they are of the form $$\al\left(\begin{array}{c}
S \\ -3V \\ -3V \\ -3S
\end{array}\right) $$
where $\al$ is a some constant. In principle one can find also this
constant by taking, for example, limit $T\rightarrow\infty$ of the
(\ref{ampl}).
\section*{Conclusions}

In the present work we give an example of non-zero correlation
functions of the type $<27,\overline{27},1_m,1_s>$, which are related
to the additional contribution to the induced mass term for Higgs
particles at low energies. We compute such correlation functions
explicitly and show that they form finite dimensional representations
of the duality group $SL_2(Z)$, acting on the untwisted moduli space.
The action of the duality group is shown to be consistent with its
action on the twisted fields characterized by the fixed points set
under the orbifold group. We have also derived the correlation function
of bosonic twist fields for the case of $g$, $g^{-1}$, $h$, $h^{-1}$
twisted sectors. Finally, we integrated the correlation functions over
2-dimensional world sheet and discuss ambiguity of the solution
using its transformation properties under the action of modular group.
\section*{Appendix}
In the presence of the twist fields $\s_{g}(z_{1})\s_{g^{-1}}(z_{2})
\s_{h}(z_{3})\s_{h^{-1}}(z_{4})$ we will need integrals of holomorpic
and antiholomorpic functions of the kind\footnote{for the case
$g=\theta^2$ and $h=\theta^3$ ($\theta=e^{2\pi i/6}$) $a=-2/3$ and
$b=-1/2$}: $$ \om(z)=(z-z_1)^a(z-z_2)^c(z-z_3)^b(z-z_4)^d $$ where
$a+c=b+d=-1$.\\
Let's denote $g=e^{2\pi ia},h=e^{2\pi ib}$.

\subsection*{A Integral over complex plane} We need to evaluate
integral $\int_{C}\om \wedge \om'$ for any two closed one-forms
$\om$ and $\om'$, such that the wedge product of which is well-
defined on complex plane (i.e. $g'=g^{-1}$,
$h'=h^{-1}$).\\
By Hodge's theorem there exists at least locally a zero-form
(function), such that $\om=df$. Using Stoke's theorem we have \be
\int_{C}\om \wedge \om'=\int_{C} df \wedge \om'=\int_{C}d(f \om')
=\int_{\p C} f \om'
\ee
The boundary is shown in Fig.\ref{boundary}. \begin{figure}
\setlength{\unitlength}{1cm}
\begin{picture}(15,4)
\put(3,4){\circle*{0.1}}\put(8,4){\circle*{0.1}}
\put(3,1){\circle*{0.1}}\put(8,1){\circle*{0.1}}
\put(3,3.4){$z_1$}\put(8.4,3.8){$z_2$}
\put(3,0.4){$z_4$}\put(8.4,0.8){$z_3$}
\put(3.2,4.1){\line(1,0){4.6}}\put(3.2,3.9){\line(1,0){4.6}}
\put(3.2,1.1){\line(1,0){4.6}}\put(3.2,0.9){\line(1,0){4.6}}
\put(7.9,1.2){\line(0,1){2.6}}\put(8.1,1.2){\line(0,1){2.6}}
\put(4,3.5){$w_2$}\put(4,4.3){$w_2'$}\put(7.4,3.5){$z_A$}\put(7.4,4.3
){$z_A'$} \put(7.4,2.7){$w_1'$}\put(8.2,2.7){$w_1$}
\put(4,1.3){$w_3$}\put(4,0.5){$w_3'$}\put(7.4,1.3){$z_B$}\put(7.4,0.5
){$z_B'$} \end{picture}
\caption{The boundary of the complex plane in the presence of twist
fields} \label{boundary}
\end{figure}
Choosing two independent loops ${\cal C}_1$ and ${\cal C}_2$, such
that $$ \oint_{{\cal C}_1}\om=(g^{-1}-1)\int_{z_1}^{z_2} \om $$ $$
\oint_{{\cal C}_1^{-1}}\om=(h^{-1}-1)\int_{z_3}^{z_4} \om $$ $$
\oint_{{\cal C}_1}\om=-\oint_{{\cal C}_1^{-1}}\om $$ \be
\oint_{{\cal C}_2}\om=(1-g)(1-h)\int_{z_2}^{z_3} \om =\g
\int_{z_2}^{z_3} \om
\ee
we have
\be
\int_{C}\om \wedge \om'=[\oint_{{\cal C}_1}+\int_b+\oint_{{\cal
C}_1^{-1}} +\int_{b^{-1}}] f \om'
\label{b_C_int}
\ee
Using the fact that quantities
\be
f(w_1)-f(w_1')=[\int_{w_1'}^{z_2}+\oint_{{\cal
C}_1}+\int_{z_2}^{w_1}]\om =\oint_{{\cal C}_1}\om
\ee
\be
gf(w_2')-f(w_2)=g[f(w_2)+\int_{w_2}^{z_1}\om (1-g^{-1})]-
f(w_2)=(g-1)f(z_1) \ee
\be
hf(w_3')-f(w_3)=h[f(w_3)+\int_{w_3}^{z_3}\om (1-h^{-1})]-
f(w_3)=(h-1)f(z_3) \ee
doesn't depend on $w$ and
$$f(z_A')=f(z_A)+\oint_{{\cal C}_1}\om$$
$$f(z_B)=f(z_B')+\oint_{{\cal C}_1^{-1}}\om$$
$$f(z_A)=f(z_B)+\int_{z_3}^{z_2} \om$$
we find
\be
[\int_b+\int_{b^{-1}}]f\om'=\int_{z_2}^{z_3}\om'(w_1)[f(w_1)-
f(w_1')]= \int_{z_2}^{z_3}\om' \oint_{{\cal C}_1}\om= {1 \over \g'}
\oint_{{\cal C}_1}\om \oint_{{\cal C}_2}\om' \label{b_int}
\ee
$$[\oint_{{\cal C}_1}+\oint_{{\cal C}_1^{-1}}]f\om'
=\int_{z_1}^{z_2}\om'(w_2)[gf(w_2')-f(w_2)] -
\int_{z_3}^{z_4}\om'(w_3)[hf(w_3')-f(w_3)]$$ $$={g' \over 1-g'}
[gf(z_A')-f(z_A)]\oint_{{\cal C}_1}\om'+ ={h' \over 1-h'} [hf(z_B')-
f(z_B)]\oint_{{\cal C}_1^{-1}}\om'$$ $$=[f(z_A)-f(z_B)+\oint_{{\cal
C}_1} \om({1 \over 1-g'}{1 \over 1-h'})] \oint_{{\cal C}_1}\om'$$
\be
=[-{1 \over \g}\oint_{{\cal C}_2}\om +\oint_{{\cal C}_1} \om ({1
\over 1-h}{1 \over 1-g})] \oint_{{\cal C}_1}\om' \label{C_int}
\ee
Substituting (\ref{b_int}) and (\ref{C_int}) into (\ref{b_C_int}) we
find \be
\int_{C} \om \wedge \om'=
{1 \over \g'} \oint_{{\cal C}_{1}} \om \oint_{{\cal C}_{2}} \om' -{1
\over \g} \oint_{{\cal C}_{1}} \om'\oint_{{\cal C}_{2}} \om +(\al-\al')
\oint_{{\cal C}_{1}} \om \oint_{{\cal C}_{1}} \om'
\label{omega_omega'}
\ee
where
$$\al={1 \over 2}\left( {1 \over 1-h}-{1 \over 1-g}\right)$$
Normalizing $\om$ and $\om'$ such that
$$ \oint_{{\cal C}_{1}} \om=1 , \oint_{{\cal C}_{1}} \om'=1 $$ and
denoting
$$ \tau=\int_{z_2}^{z_3} \om-\al,\,\tau'=\int_{z_2}^{z_3} \om'-\al'$$
we get
\be
\int_{C} \om \wedge \om'=\tau'-\tau
\ee
Consequently, for
$$ \om_1=z^a(z-x)^c(z-1)^b$$
$$ \om_2=z^c(z-x)^a(z-1)^{-1-b}$$
where a+c=-1, we have
$$\int_{C} \om_1 \wedge \bar \om_1=\bar \tau-\tau$$ $$\int_{C}
\om_2 \wedge \bar \om_2=\bar \tau-\tau$$ since
$$0=\int_{C} \om_1 \wedge \om_2=\tau_2-\tau_1$$

\subsection*{B Contour integrals}
If we denote
$$ \om(a,c,b)=z^a(z-x)^c(z-1)^b$$
then
$$\oint_{{\cal C}_1} \om(a,c,b)=(g^{-1}-1) \int_0^x \om(a,c,b) =(g^{-
1}-1)e^{-i\pi (b+c)} I_1(a,c,b)$$
$$\oint_{{\cal C}_2} \om(a,c,b)=\g \int_x^1 \om(a,c,b) =\g e^{-i\pi b}
I_2(a,c,b)$$
where
$$I_1(a,c,b) \equiv \int_0^x z^a(x-z)^c(1-z)^b$$
$$=x^{1+a+c}{\Gamma(a+1)\Gamma(c+1) \over \Gamma(a+c+2)}F(a+1,-
b;a+c+2;x)$$ $$I_2(a,c,b) \equiv \int_x^1 z^a(z-x)^c(1-z)^b$$ $$=(1-
x)^{1+c+b}
{\Gamma(b+1)\Gamma(c+1) \over \Gamma(c+b+2)}F(b+1,-a;c+b+2;1-
x)$$ (Here $F$ is hypergeometric function.)\\ Using $g=e^{2\pi i
a},h=e^{2\pi i b},\g=(1-g)(1-h)$ we find $$\oint_{{\cal C}_1}
\om(a,c,b)=-2ie^{-i\pi(a+c+b)}s(a)x^{1+a+c}$$
$$\times{\Gamma(a+1)\Gamma(c+1) \over \Gamma(a+c+2)}F(a+1,-
b;a+c+2;x)$$ $$\oint_{{\cal C}_2} \om(a,c,b)=-4e^{i\pi a}s(a)s(b)(1-
x)^{1+c+b}$$ \be
\times{\Gamma(b+1)\Gamma(c+1) \over \Gamma(c+b+2)}F(b+1,-
a;c+b+2;1-x) \label{con_int}
\ee
where $s(a)=\sin(\pi a)$. In the case $a+c=-1$ (\ref{con_int})
becomes $$\oint_{{\cal C}_1} \om(a,-1-a,b)=-2ie^{-i\pi b} \pi F(a+1,-
b;1;x)$$ $$\oint_{{\cal C}_2} \om(a,-1-a,b)=-4e^{i\pi a}s(a)s(b)(1-
x)^{b-a}$$ $$\times{\Gamma(b+1)\Gamma(-a) \over \Gamma(b-
a+1)}F(b+1,-a;b-a+1;1-x)$$ {\bf Acknowledgements}

I would like to express the gratitude and my special thanks to Prof.
K.Narain for his help and useful discussions.

This work was supported by the International Centre for Theoretical
Physics, Trieste, the International Atomic Energy Agency, and the
United Nations Educational, Scientific and Cultural Organization.

\end{document}